\documentstyle[aps,preprint,epsfig]{revtex}
\begin{document}
\title
{SCALE FACTOR IN DOUBLE PARTON COLLISIONS AND PARTON DENSITIES IN TRANSVERSE
SPACE} 

\author{A. Del Fabbro and D.Treleani} 

\address{ Universit\`a di Trieste; Dipartimento di Fisica Teorica, 
Strada Costiera 11, Miramare-Grignano, and
INFN, Sezione di Trieste, I-34014 Trieste, Italy.}

\maketitle

\begin{abstract}
The scale factor
$\sigma_{eff}$, which characterizes double parton collisions in
high energy 
hadron interactions, is a direct manifestation of the distribution of the
interacting partons in
transverse space, in such a way that different distributions
give rise to different values of $\sigma_{eff}$
in different double parton collision processes.
We work out the value of the scale factor
in a few reactions of interest, in a correlated
model of the multi-parton density of the proton recently proposed.
\end{abstract}
\vspace{3cm}

E-mail delfabbr@trieste.infn.it \\

E-mail daniel@trieste.infn.it \\

\newpage

\section{Introduction}
Double parton collisions are a new feature of high energy hadron interactions
which becomes increasingly important at high energies. The number of partons that
can undergo a hard collision is in fact a fast growing function of $s$
and, as a consequence, when the c.m. energy 
is large enough, the probability of having more than one hard partonic 
interaction in
the same inelastic event becomes sizable. Double parton collisions, 
foreseen long ago by several authors\cite{dpth}, have been
in fact observed recently by CDF\cite{cdf}. The non-perturbative input to 
the double parton collisions 
are the double parton distributions, which are independent on the
parton distributions usually considered in large$-p_t$ physics, since they
are related directly to the two-body
parton correlation of the hadron structure. In the simplified hypothesis of neglecting all correlations in fractional
momenta, the inclusive double parton 
scattering cross
section for the two parton processes $A$ and $B$ reduces nevertheless to 
a simplest
factorized expression:
\begin{equation}
\sigma_D=m\frac{\sigma_A \sigma_B}{2\sigma_{eff}}
\label{fact}
\end{equation}
where $m=1$ when $A$ and $B$ are indistinguishable processes and $m=2$ when they
are distinguishable. $\sigma_A$ and $\sigma_B$ are the inclusive single scattering cross sections
for producing the processes $A$ and $B$ respectively  and all the new information on the
structure of the hadron in transverse space is summarized in the value of the
scale 
factor $\sigma_{eff}$. This simplest hypothesis has not been contradicted by
the experimental evidence\cite{cdf}, whose results are in fact described with a
single parameter (the scale factor $\sigma_{eff}$). The experimental value quoted
by CDF,
$\sigma_{eff}=14.5\pm1.7^{+1.7}_{-2.3}$ mb, is however too small to be
understood in a simplest uncorrelated picture of the multi-parton distribution
and indicates that correlations in transverse
space play an important role\cite{Calucci:1998ii}. A possibility which has been
considered\cite{Calucci:1999yz} to explain the smallness of the
value of $\sigma_{eff}$ is to correlate the population of gluons and sea
quarks with the configuration
of the valence in transverse space, in such a way that the average number of
gluons and sea quarks is small when the
valence quarks are all close in transverse space and, on the contrary, is
large when they are separated by a (relatively) large transverse 
distance. Such a mechanism
increases the dispersion in the 
number of multiple parton collisions and, as a consequence, 
$\sigma_D$ (which is 
proportional to that dispersion\cite{Calucci:1999yz}). 
The value of
$\sigma_{eff}$ is therefore diminished with respect to the uncorrelated case.
The scale factor, on the other hand, is the result of the
overlap in transverse space of the matter distribution of the two interacting
hadrons and
a feature of the model above is that the average transverse
distance of a pair of valence quarks is different as
compared with the average transverse distance of a pair
of gluons or sea quarks. In the model 
$\sigma_{eff}$ is therefore different for
double parton scatterings involving valence quarks and for double
parton scattering involving sea quarks and gluons. Although the
details of the model should be understood
in a qualitative rather than in a quantitative sense, it is rather natural 
to expect different values of the scale factor in different
reactions. We think therefore that it may be
interesting to have an indication, even if only qualitative, on 
the size of the effect and, to that purpose, we work out in the present
note the values of the scale factor foreseen in the model \cite{Calucci:1999yz}
in a few cases of interest.

\section{Scale factor in a correlated model}

At large energies one will be able to observe double parton
collisions in various
channels\cite{Catani:2000jh}\cite{DelFabbro:2000tf}\cite{stirling}. 
In particular the production of
two equal sign $W$ bosons, at relatively low $p_t$, goes almost entirely 
through double parton
collisions\cite{stirling} and is originated, to a large extent, by 
interactions with 
valence quarks. Another case where double
parton collisions are expected to play an important role is in the
production of two $b\bar b$ pairs. The corresponding integrated cross section 
is in fact rather large, by using Eq.\ref{fact} one may estimate that
the double parton scattering cross section for producing two $b\bar b$ pairs is
of the order of $10\mu{\rm b}$ at the energy of the LHC. 
The dominant mechanism of $b{\bar b}$ production is
gluon fusion and the observation of double parton
collisions, in a equal sign $W$ boson pair production and in the production of two
$b\bar b$ pairs, allows therefore one to compare the
distribution in transverse space of valence quarks with the distribution in
transverse space of gluons. 

To work out the scale factor in the two cases we write the inclusive double 
parton scattering cross section, for the parton
interactions $A$ and $B$, as
\begin{equation}
\sigma_D(A,B)=\frac{m}{2}\sum_{ijkl}\int_{p_t^{cut}}\Gamma_{ij}(x_1,x_2;b)
\hat{\sigma}_{ik}^A(x_1,x_1')
\hat{\sigma}_{jl}^B(x_2,x_2')\Gamma_{kl}'(x_1',x_2';b)dx_1dx_1'dx_2dx_2'd^2b \,
\label{sigmad2}
\end{equation}
where $\Gamma_{ij}(x_1,x_2;b)$ are the double parton 
 distributions, depending on the two fractional momenta of the partons
$x_1$ and $x_2$ and on their relative transverse distance $b$. 
The indices $i$ and $j$ refer to the different kinds of
 partons, $\hat{\sigma}_{ik}^A$ and $\hat{\sigma}_{jl}^B$ are the partonic cross
 sections and the QCD dependence on the scale of the interaction is implicit in
 all quantities. In the case we are considering the
dependence of $\Gamma$ on $x_1$, $x_2$ and $b$ is factorized:
\begin{equation}
\Gamma_{ij}(x_1,x_2;b)=G_i(x_1)G_j(x_2)F_j^i(b)
\label{gfact}
\end{equation}
where $G_i(x)$ are the usual parton distributions.
If $F_j^i(b)$ do not depend on $i$ and $j$ one obtains Eq.\ref{fact} and
the scale factor $\sigma_{eff}$ is universal.
In general the double scattering
cross section is however written as
\begin{equation}
\sigma_D(A,B)=\frac{m}{2}\sum_{ijkl}\Theta^{ij}_{kl}\sigma_{ij}(A)\sigma_{kl}(B)
\label{sigmad1}
\end{equation}
where $\sigma_{ij}(A)$ is the hadronic inclusive cross section for two partons of kind
$i$ and $j$ to undergo the hard interaction $A$, and
\begin{equation}
\Theta^{ij}_{kl}=\int d^2bF_k^i(b){F_l^j}'(b)
\label{fb}
\end{equation}
are geometrical coefficients, with dimensions of the inverse of a cross section, depending
on the various parton processes.

In the model \cite{Calucci:1999yz} the probability to find the proton in a
given configuration, with $n$ gluons or sea-quarks, has the following 
expression:

\begin{eqnarray}
P(X_v,{\bf B}_v;x_1,{\bf b}_1,\dots x_n,{\bf
b}_n)&&=q_v(X_1)q_v(X_2)q_v(X_3)\varphi({\bf B}_D,{\bf B})\nonumber\\
\times\frac{1}{n!}
\bigg[g(x_1)f( B_D, b_1)&&
\dots g(x_n)f( B_D, b_n)\bigg]
{\rm exp}\Bigg\{-\frac{B_D^2}{\langle B_D^2\rangle}\int g(x)dx \Bigg\}
\label{pdistr2}
\end{eqnarray}
where $q_v(X)$ are the inclusive distributions of valence quarks, 
as a function of the momentum fraction $X$, and $\varphi({\bf
B}_D,{\bf B})$ is the density of valence quarks in transverse space, 
where the coordinates of valence quarks 
in transverse space ${\bf B}_v$ are given by 
\begin{eqnarray}
{\bf B}_1&=&\frac{1}{2}{\bf B}_D+{\bf B}\nonumber\\
{\bf B}_2&=&\frac{1}{2}{\bf B}_D-{\bf B}\nonumber\\
{\bf B}_3&=&-{\bf B}_D
\end{eqnarray}
In the model $\varphi({\bf B}_D,{\bf B})=\int dZ_DdZ\phi({\bf R}_D,{\bf R})$,
where $\phi({\bf R}_D,{\bf R})$ represents the distribution density
of the proton in coordinate space. The explicit expression
used is
\begin{equation}
\phi({\bf R}_D,{\bf R})=\frac{\lambda_D^3\lambda^3}{(8\pi)^2}
\exp\big\{-(\lambda_DR_D+\lambda R)\big\} 
\end{equation}
where
\begin{eqnarray}
\lambda_D&=&\frac{2\sqrt{3}}{\sqrt{\langle r^2\rangle}}\nonumber\\
\lambda&=&\frac{4}{\sqrt{\langle r^2\rangle}}
\end{eqnarray}
with $\sqrt{\langle r^2\rangle}=0.81$fm, the proton charge radius.

Given a configuration of the valence, the
average density of gluons and sea quarks in a point with transverse coordinate 
$b$ and momentum fraction $x$ is given by $g(x)f( B_D, b)$. 
The overall average number of gluons and sea quarks at a given $x$ (namely
after integrating on $b$ and on the configurations of the valence) is $g(x)$ and is identified
 with the inclusive distribution of gluons and sea quarks. In the same way the
inclusive distributions of the valence quarks, $q_v(X)$, are the result of
integrating over the transverse coordinates ${\bf B}_v$ and of summing over all configurations of gluons and sea quarks. 

The dependence of the average density of gluons and sea quarks on the 
transverse coordinate $b$ is expressed by
\begin{equation}
f( B_D, b)=\frac{3}{2\pi}\Biggl(1-\frac{b^2}{B^2_D}\Biggr)^{1/2}\frac{1}
{\langle B_D^2\rangle}\theta(B_D-b)
\label{f}
\end{equation}
which is the projection on the transverse plane of a sphere of radius 
$B_D$, rescaled with the factor 
$B_D^2/\langle B_D^2\rangle$, where the average $\langle B_D^2\rangle$ is taken
with the density $\varphi({\bf
B}_D,{\bf B})$. The average density of gluons and sea quarks and
the corresponding average number (which grows as
$B_D^2/\langle B_D^2\rangle$) are therefore correlated with
the configuration of the valence, while
all fractional momenta are uncorrelated. 

The model distinguishes two different kinds of
partons, as far as the number and density in transverse space are concerned,
the valence quarks and the sea quark and gluons. The indices $i$ and $j$,
as a consequence, can take two different values, $v$ and $s$, and 
the relevant transverse densities of parton pairs to be considered are 
\begin{eqnarray}
F_v^v(b)&&=\frac{1}{6}\sum_{{j\neq i}=1}^{3}\int d^2Bd^2B_D\varphi({\bf
B}_D,{\bf B})\delta({\bf B}_i-{\bf B}_j-{\bf b})\nonumber\\
F_s^v(b)&&=\frac{1}{3}\sum_{i=1}^{3}\int d^2Bd^2B_Dd^2b'\varphi({\bf
B}_D,{\bf B})f( B_D, b')\delta({\bf B}_i-{\bf b}'-{\bf b})\nonumber\\
F_s^s(b)&&=\int d^2Bd^2B_Dd^2b'd^2b''\varphi({\bf
B}_D,{\bf B})f( B_D, b')f(
B_D, b'')\delta({\bf b}'-{\bf b}''-{\bf b})
\end{eqnarray}
By using Eq.\ref{fb} and the expression below, the matrix $\Theta^{ij}_{kl}$
and the effective cross section are 
readily evaluated:
\begin{equation}
\sum_{ijkl}\Theta^{ij}_{kl}\sigma_{ij}(A)\sigma_{kl}(B)=
\frac{\sigma(A)\sigma(B)}{\sigma_{eff}}
\label{seff}
\end{equation}

\section{Results}

The resulting values of
$1/\Theta^{ij}_{kl}$ are shown in the table. 
The scale factor $\sigma_{eff}$ is plotted as a
function of the c.m. energy in
fig.1 and in fig.2, corresponding to $pp$ and $p{\bar p}$ interactions 
respectively, in various processes of interest: $i$) production of two equal
sign $W$ bosons, $ii$) production of a $W$ boson of either positive or
negative sign together with two jets or $iii$) with a $b{\bar b}$ pair,
$iv$) production of four jets, $v$) production of two $b{\bar b}$
pairs. In the case of jets the scale factor has been evaluated by using as a
lower cutoff in $p_t$ the value of $5GeV$ (the scale factor is however rather
insensitive to that choice). All the single scattering cross sections in
Eq.\ref{seff} have been computed at the lowest order in the coupling constant and
by making use of the Martin-Roberts-Stirling 1999 (MRS99) parton 
distributions\cite{mrs99}. The different results obtained for $W^+W^+$ and
$W^-W^-$ production for $pp$ collisions is mainly due to the different content
of $d$ and $u$ quarks in the proton. A consequence is in fact the different 
contribution of the sea quarks in the two cases,
whose distribution is sizably different in the model in
comparison with the distribution in transverse space of the valence. 
To check the dependence on the choice
of the distribution functions 
we have repeated the calculation by using the CTEQ5 parton
distributions\cite{cteq00}. In all cases the scale factor is not 
changed by more than a few percent. In fact the results for $\sigma_{eff}$ 
are rather insensitive to the choice of parton distributions, 
since $\sigma_{eff}$ is obtained by making ration of cross sections.
The main qualitative feature of the results obtained is the 
strong difference, at Tevatron energy, between final states
with and without a $W$ boson, and the energy dependence of the scale factor, 
which
is sizable when moving from Tevatron to LHC energy in the  
channels containing a $W$ boson. 

While the actual values should be
considered only as indicative, the qualitative features just pointed out are
likely to be of more general validity, since they are originated by the
different source of initial state
partons in the case of $W$ production, which in most cases involves 
valence quarks,
in comparison with the other channels
considered, that are mainly generated by interactions of 
sea quarks and gluons. The overall indication  
is that it is plausible to expect non minor differences in the
values of $\sigma_{eff}$ in different processes.
It is also apparent that a basic element, to understand
the three dimensional structure of the proton, is a
quantitative information on the  
correlations in transverse space of the various pairs of initial state partons 
and that the
scale factors, of the different double parton collision processes, are the
physical observables which can provide that information.

\vskip.25in
{\bf Acknowledgment}
\vskip.15in
This work was partially supported by the Italian Ministry of University and of
Scientific and Technological Researches (MURST) by the Grant
COFIN99.

	\vskip.25in
	\begin{table}
	 \caption{Scale factors $1/\Theta^{ij}_{kl}$ (mb)}
 	\label{tab1}
  	\begin{tabular}{ccccc}
   	 Partons(ik-jl)         &Scale factor\\
   	\tableline
	(ss-ss)     & 12.4  \\
	(vs-ss)     & 31.9  \\  	
	(vv-ss)     & 28.3  \\  
	(vs-vs)     & 69.4  \\  
	(vs-sv)     & 69.4  \\    
	(vv-vs)     & 68.3  \\  
	(vv-vv)     & 67.4  \\  
	\end{tabular}
	\end{table}	
	\vskip.25in

\begin{figure}
\vspace{4 cm}
\centerline{
\epsfysize=10cm \epsfbox{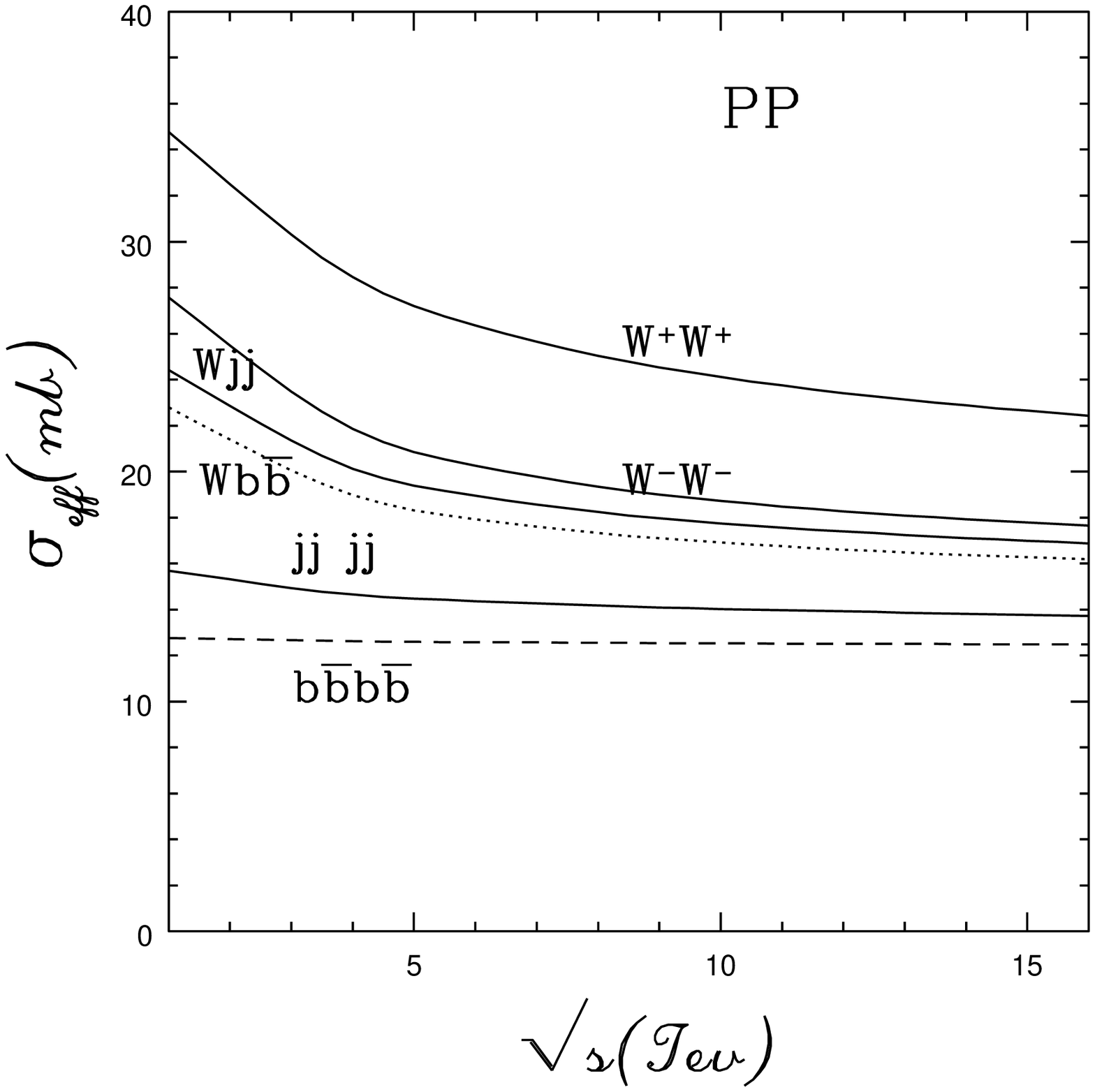}}
\vspace{2 cm}
\caption[ ]{$\sigma_{eff}$ as a function of the c.m. energy in various
processes  in $pp$ collisions.}
\label{lhc}
\end{figure} 

\begin{figure}
\vspace{4 cm}
\centerline{
\epsfysize=10cm \epsfbox{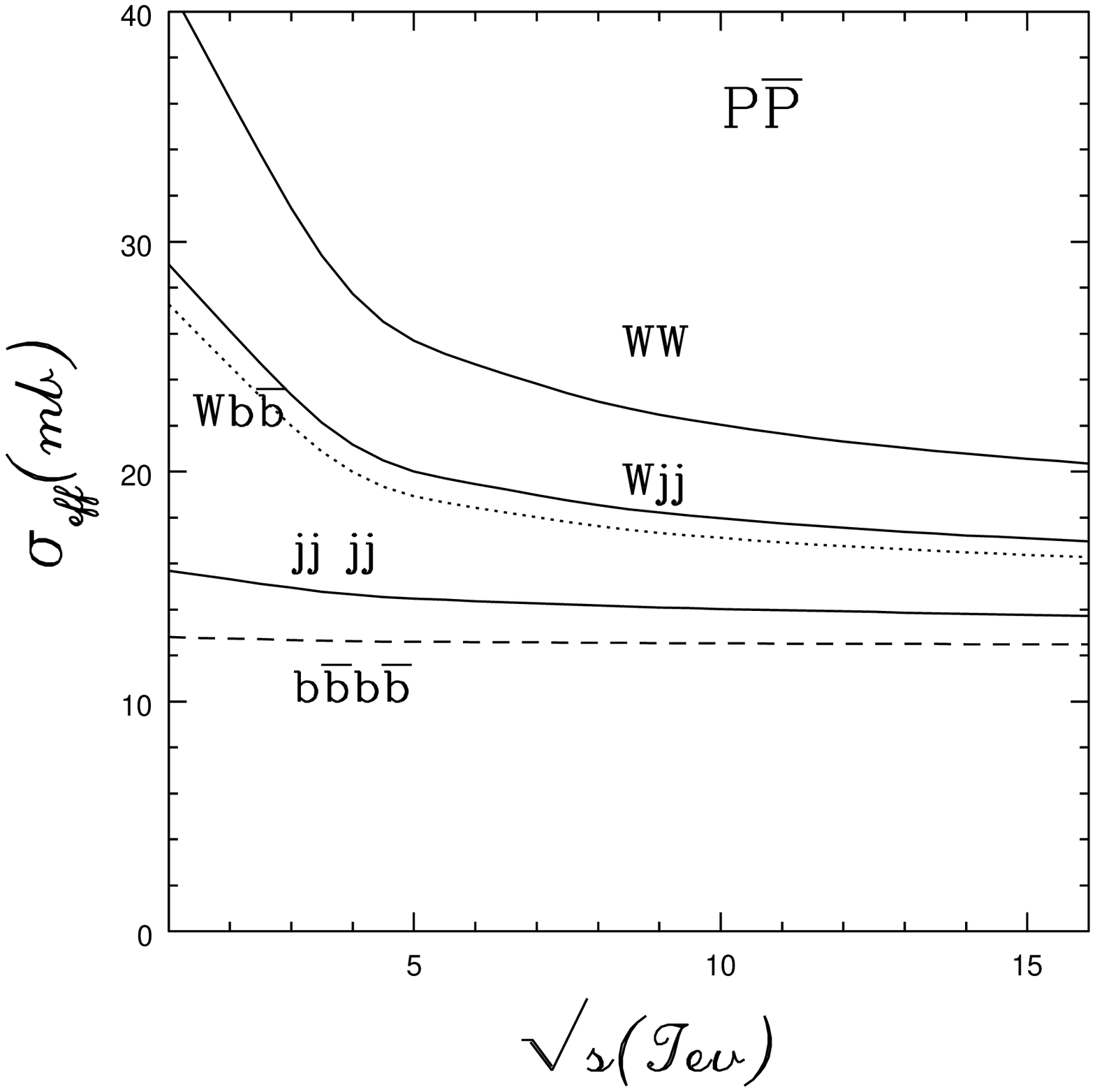}}
\vspace{2 cm}
\caption[ ]{$\sigma_{eff}$ as a function of the c.m. energy in various
processes in $p{\bar p}$ collisions.}
\label{tev}
\end{figure}

\end{document}